\documentclass[runningheads]{llncs}
\usepackage{times}
\usepackage{latexsym}
\usepackage{graphicx}
\usepackage{hyperref}
\usepackage{url}
\usepackage{cite}
\usepackage{amsmath,amsfonts,amssymb,amscd}
\usepackage{float}
\usepackage{subcaption}
\usepackage{tikz}
\usetikzlibrary{calc}
\usetikzlibrary{shapes}
\pgfdeclarelayer{bg}    
\pgfsetlayers{bg,main} 

\title{New Datasets and a Benchmark of Document Network Embedding Methods for Scientific Expert Finding}
\titlerunning{Datasets and Benchmark of DNE Methods for Expert Finding}

\author{Robin Brochier\inst{1} 
\and
 Antoine Gourru\inst{1}  \and
  Adrien Guille\inst{1} \and
  Julien Velcin\inst{1} }
  \authorrunning{R. Brochier et al.}
\institute{ Universit\'e de Lyon, Lyon 2 \\
  ERIC EA3083\\\email{\{firstname\}.\{lastname\}@univ-lyon2.fr}}

\begin{document}
\bibliographystyle{splncs04}
\maketitle

{\let\thefootnote\relax\footnotetext{Copyright \textcopyright\ 2020 for this paper by its authors. Use permitted under Creative Commons License Attribution 4.0 International (CC BY 4.0). BIR~2020, 14 April 2020, Lisbon, Portugal.}}

\begin{abstract}
The scientific literature is growing faster than ever. Finding an expert in a particular scientific domain has never been as hard as today because of the increasing amount of publications and because of the ever growing diversity of expertise fields. To tackle this challenge, automatic expert finding algorithms rely on the vast scientific heterogeneous network to match textual queries with potential expert candidates.    
In this direction, document network embedding methods seem to be an ideal choice for building representations of the scientific literature. Citation and authorship links contain major complementary information to the textual content of the publications.
In this paper, we propose a benchmark for expert finding in document networks by leveraging data extracted from a scientific citation network and three scientific \textit{question \& answer} websites. We compare the performances of several algorithms on these different sources of data and further study the applicability of embedding methods on an expert finding task.            
\end{abstract}

\section{Introduction}

Many tools offer to search and filter the vast data sources available on the Web. In particular, there is a multitude of platforms directed to the scientific community. From the simple search engine for publications to the social network for researchers, all consume and produce valuable data for searching scientific content of interest. Expert finding is one the the most challenging problem that finds application in both academia and the industry. To tackle this challenge, recent advances in document network embedding (DNE) has the potential to inspire new unsupervised models that can deal with the heterogeneous network of documents of the scientific literature. However, the design of such efficient algorithms heavily depends on the development of strong evaluation frameworks.

In this paper, we propose a methodology and provide 4 datasets that extend the limited scope of expertise retrieval evaluation frameworks. Furthermore, we provide experiment results computed with unsupervised methods and we extend document network embedding algorithms to this specific task.   

Our contributions are the following:

\begin{itemize}
    \item we provide 4 datasets for expert finding extracted from a scientific publication network and three \textit{question \& answer} (\textit{Q\&{}A}) websites and make them publicly available \footnote{\url{https://github.com/brochier/expert_finding}};
    \item we describe an evaluation methodology based on the ranking of expert candidates given a set of labeled document queries; 
    \item we report experiment results that give some insights on this expert finding task;   
    \item we explore and analyze the use of state-of-the-art document network embedding algorithms for expert finding and we show that further research is needed to bridge the gap between DNE methods and expert finding. 
\end{itemize}

The rest of the paper is organized as follows. In Section \ref{sec:ref}, we survey related works. We detail in Section \ref{sec:eval} our evaluation methodology, the datasets we extracted, the evaluation measures and the algorithms we use. In Section \ref{sec:xp} we show and analyze the results of our experiments. Finally, in Section \ref{sec:conclu}, we discuss our findings and provide future directions.  

\section{Related Works} \label{sec:ref}

In this section, we first present a formal definition for expert finding. Then we present algorithms of the literature that address expert finding. Finally, we describe recent methods for document network embedding that have the potential to deal with this particular task.  

\subsection{Formal definition of expert finding}

The concept of expert finding can cover a large range of tasks. The main principle behind expertise retrieval is the search for candidates given a query. To match these two, an algorithm will be provided with some data to link the output space, a ranking of candidates, with the input space, which is often a textual content. However, many different types of data can be considered to address this challenge. To fairly compare algorithms, we choose a fixed structure for the data which reflects common use cases.   
Furthermore, if supervised methods benefit from labeled fields of expertise associated with the candidates, they are beyond the scope of this paper which focuses on unsupervised methods only. Our goal is to compare methods that do not require sometimes costly annotations.   

Early works in expert search \cite{craswell2005overview} usually consider a small set of topical queries. The direct namings of these topics are used to retrieve a list of candidates by leveraging a collection of documents they published (e.g., emails, scientific papers). This type of evaluation is used across several public datasets \cite{zhang2007expert, macdonald2007overview, mislevy2011evidence}. 

More recently, the concept of expert finding has been merged into the wider concept of entity retrieval \cite{balog2010overview}. As more and more complex data are produced on the Web, expert finding becomes a particular application of entity search.
At the same time, \textit{Q\&A} websites such as Stack Overflow \footnote{\url{https://stackoverflow.com/}} generate and make publicly accessible a big amount of questions with expert answers, collaboratively curated by their users. Several works address the search for experts in such websites \cite{riahi2012finding, zhao2014expert}. Often, the task consists in either finding the exact list of users who answered a specific question or ranking the answers according to the user votes. In the first case, the task involves considering the evolution of the users across time and, in the second case, the task involves understanding the intrinsic quality of a written answer. Nevertheless, \cite{yuan2020expert} reviews several models for expert finding in \textit{Q\&A} websites. Their experiments show that matrix factorization-based methods perform better than tree based and ranking based methods.

In this paper, we adopt the document-query methodology recently proposed in \cite{brochier2018impact}. The expert search is performed given a set of queries that are particular textual instances of some expert topics (or fields of expertise). Given a query, an algorithm should rank first the candidates that are associated to the same fields of expertise. We provide 4 datasets for which we annotated experts and document queries. Each dataset consists in candidates and documents linked by authorship relations (candidate-document e.g. authorship) and by response relations (document-document e.g citation or answer). 
A query is therefore one of the documents (e.g. a scientific paper or a question) for which we aim to retrieve some experts of the topics depicted in it. This configuration reflects many real case scenarios such as (1) the automatic search for scientific reviewers, (2) the recommendation of expert users in \textit{Q\&A} websites or even (3) the retrieval of interesting profiles for job offers.         

\subsection{Algorithms for expert finding}

Numerous works have addressed automatic expertise retrieval. We describe here the main approaches and some interesting recent methods.   
P@noptic Expert \cite{craswell2001p} creates meta-documents for a candidate by concatenating the contents of all documents she produced. In this manner, ranking the candidates given a query becomes a similarity search between the query representation and the meta-documents representations. 
A voting model \cite{macdonald2006voting} computes the similarities between the query and the documents. The algorithm then aggregates these scores at the candidate level by using a fusion technique such as the reciprocal rank \cite{zhang2003expansion}. 
A propagation model \cite{serdyukov2008modeling} takes advantage of the links between candidates and documents to propagate the similarities between the query and the documents. Using random walks with restart \cite{page1999pagerank}, the iterative propagation of the scores converges in a few steps to a stationary distribution over the candidates.
WISER \cite{Cifariello2019WISERAS} models each candidate as a small, weighted, sub-graph of the Wikipedia Knowledge Graph. Information derived from these graphs and traditional document retrieval techniques are combined to identify experts w.r.t a query. Note that methods leveraging external data are out of the scope of our benchmark.
LT Expertfinder is an evaluation framework for expert finding \cite{fischer2019lt} based on an interactive tool. It integrates various existing algorithms (such as \cite{balog2010overview}) in a user-friendly way. The underlying corpus used by this tool is the ACL Anthology Network. However, it does not include a well-established ground truth to assess who are the experts. Indeed, the evaluation is purely done in an online manner since the user has to evaluate the degree of expertise based on several features, such as author's citations, h-index, keywords, etc.
Recent works \cite{VanGysel2016experts, dargahi2017skill} propose ad hoc embedding techniques, whereas, in this work, we're interested in measuring the performance of conventional network embedding techniques.


\subsection{Document network embedding}

Network embedding \cite{perozzi2014deepwalk, grover2016node2vec} provides an efficient approach to represent nodes in a low dimensional vector space, suitable for solving various machine learning tasks. Recent techniques extend NE for document networks.
Text-Associated DeepWalk (TADW) \cite{yang2015network} extends DeepWalk to deal with textual attributes. Yang \textit{et al.} prove, following the work in \cite{levy2014neural}, that Skip-Gram with hierarchical softmax can be equivalently formulated as a matrix factorization problem. TADW then consists in constraining the factorization problem with a pre-computed representation of the documents by using Latent Semantic Analysis (LSA) \cite{deerwester1990ilsa}. 
Graph2Gauss (G2G) \cite{bojchevski2018deep} is an approach that embeds each node as a Gaussian distribution instead of a vector. The algorithm is trained by passing node attributes through a non-linear transformation via a deep neural network (encoder).      
GVNR-t \cite{brochier2019global} is a matrix factorization approach for document network embedding, inspired by GloVe \cite{pennington2014glove}, that simultaneously learns word, node and document representations by optimizing a least-square objective over a co-occurrence matrix of the nodes constructed by truncated random walks. 
IDNE \cite{brochier2020inductive} introduces a Topic-Word Attention mechanism, trained from the connections of a document network, to represent documents as mixtures of topics. 

DNE algorithms do not directly apply to expert finding data since they are not designed to handle multiple types of nodes, in particular candidate nodes. In this paper, we show (1) two methods to extend their applicability to the task of expert finding and (2) the impact of their representations when they are used as document representations for traditional expert finding algorithms. 

\section{Evaluation Methodology} \label{sec:eval}

We present in this section the evaluation methodology that we follow to access the performances of several algorithms for expert finding. We first describe the task we seek to solve, then we describe the datasets that we extracted and explain how we annotated them in order to access the quality of the algorithms' outputs. Finally, we detail the models used in our experiments.   

\subsection{Ranking expert candidates from document queries}

Expert finding is a complex task that can be formalized in multiple ways. Early works define this task as a ranking problem given several topic-queries where the naming of these topics are directly used as queries to retrieve the expert candidates. However, in many real world applications, a user is asked to provide a specific and detailed query. 
In a \textit{Q\&{}A} website for instance, a user usually exposes the problem she faces in full detail and does not necessarily know the exact naming of the fields of expertise needed to solve her problem. Furthermore, querying an algorithm with a small set of topic-queries can lead to poor evaluation measures due to the usually small number of fields of expertise associated with the dataset. For this reasons, we follow the document-query evaluation methodology proposed in \cite{brochier2018impact} by processing 4 datasets for which a set of document-queries is manually annotated. 

The expert finding task in this paper is a ranking problem. Given a document labeled with a ground truth set of fields of expertise, an algorithm is queried to rank a set of candidates, among which a subset of experts are associated with the same set of labels. The data provided to the algorithms consists in a corpus of $n_d$ documents $D$, $n_c$ candidates $C$, a network of authorship with adjacency matrix $A_{dc} \in \mathbb{N}^{n_d \times n_c}$ and a network of documents with adjacency matrix $A_{dd} \in \mathbb{N}^{n_d \times n_d}$. Figure \ref{fig:data} shows an hypothetical dataset used in this paper.
The ranking is performed in an unsupervised setting, that is, no ground truth labels of expertise are given to the algorithms. The set of labeled documents (the queries) can be smaller than $n_d$ and the set of labeled candidates (experts) can be smaller than $n_c$ (i.e not all documents and candidates are labeled). 

\tikzset{
   pics/paper/.style args={#1}{
       code={
        \node[transform shape, circle, minimum size=2cm, opacity=1] (#1) at (0,0) {};
        \draw (-0.4,-0.6) -- (-0.4,0.6) -- (0.4,0.6) -- (0.4,-0.6) -- (-0.4,-0.6);
        \draw (-0.3,0.3) -- (0.3,0.3);
        \draw (-0.3,-0.3) -- (0.3,-0.3);
        }
   }
}

\tikzset{
   pics/author/.style args={#1}{
       code={
        \node[transform shape, circle, minimum size=2cm, opacity=1] (#1) at (0,0) {};
        \draw (0,0.5) circle[radius=0.35];
        \draw (0.65,-0.5) arc(0:180:0.65) -- cycle;
        }
   }
}

\begin{figure}[ht]
\centering
\begin{tikzpicture}[scale=0.45]
    \foreach \x in {1,...,6}{
        \pic at (\x *24/7,4) {paper=p\x};
        \node[] at (\x *24/7,4) {$D_{\x}$};
    }
    \foreach \x in {1,...,5}{
        \pic at (\x *24/6,10) {author=a\x};
        \node[] at (\x *24/6,9.5) {$C_{\x}$};
    }
    \node[scale=2] at (0,7) {$A_{dc} \text{~~} \Big\{ $};
    \node[scale=2] at (0,1) {$A_{dd} \text{~~} \Big\{ $};
    \node[scale=2] at (0,4) {$D \text{~~~~~~} \Big\{ $};
    \node[scale=2] at (0,10) {$C \text{~~~~~} \Big\{ $};
        
    \coordinate (shift) at (1.2,0);
    \coordinate (dshift) at (1.2,1);
    \node[draw, fill=gray!80, star, scale=0.7](s) [right] at ($(a3)+(shift)$) {};
    \node[draw, fill=gray!80, star, scale=0.7] [right] at ($(a4)+(shift)$) {};
    \node[draw, fill=gray!80, ellipse, scale=0.7](ee) [right] at ($(a4)+(dshift)$) {};
    \node[draw, fill=gray!80, ellipse, scale=0.7](e) [right] at ($(a5)+(shift)$) {};
    
    \node[draw, fill=gray!80, star, scale=0.7] [right] at ($(p1)+(shift)$) {};
    \node[draw, fill=gray!80, star, scale=0.7] [right] at ($(p3)+(shift)$) {};
    \node[draw, fill=gray!80, star, scale=0.7] [right] at ($(p5)+(shift)$) {};
    \node[draw, fill=gray!80, ellipse, scale=0.7] [right] at ($(p5)+(dshift)$) {};
    \node[draw, fill=gray!80, ellipse, scale=0.7](eee) [right] at ($(p6)+(shift)$) {};
    
    \node[scale=1](lab) at (15,14) {Expertise labels};
    \draw[black!40, line width=0.15mm, dashed, ->] (ee) to[out=90,in=-60] (lab);
    \draw[black!40, line width=0.15mm, dashed, ->] (s) to[out=90,in=-90] (lab);

    \draw[black!80, line width=0.35mm, -] (a1) -- (p1);\draw[black!80, line width=0.35mm, -] (a1) -- (p2);\draw[black!80, line width=0.35mm, -] (a1) -- (p3);
    \draw[black!80, line width=0.35mm, -] (a2) -- (p2);
    \draw[black!80, line width=0.35mm, -] (a3) -- (p3);\draw[black!80, line width=0.35mm, -] (a3) -- (p4);\draw[black!80, line width=0.35mm, -] (a3) -- (p5);
    \draw[black!80, line width=0.35mm, -] (a4) -- (p5);\draw[black!80, line width=0.35mm, -] (a4) -- (p6);
    \draw[black!80, line width=0.35mm, -] (a5) -- (p6);
    
    \begin{pgfonlayer}{bg}    
     \draw [black!80, line width=0.35mm, dotted, -] (p1) to[out=-90,in=-90] (p2);
     \draw [black!80, line width=0.35mm, dotted, -] (p1) to[out=-90,in=-90] (p4);
     \draw [black!80, line width=0.35mm, dotted, -] (p3) to[out=-90,in=-90] (p5);
     \draw [black!80, line width=0.35mm, dotted, -] (p2) to[out=-90,in=-90] (p3);
     \draw [black!80, line width=0.35mm, dotted, -] (p5) to[out=-90,in=-90] (p6);
     \draw [black!80, line width=0.35mm, dotted, -] (p4) to[out=-90,in=-90] (p6);
    \end{pgfonlayer}
\end{tikzpicture}
\caption{Hypothetical example of an expert finding dataset we use in this paper. 5 candidates are authors of 6 documents. The 6 documents are connected to each other by citation in a scientific corpus, or by answer in a same post in a \textit{Q\&{}A} website. Among the candidates, 3 are known to be experts in stars and/or in circles. 4 documents are associated to these 2 fields of expertise as well. In our evaluation methodology, we query an algorithm with these 4 documents and expect a ranking of candidates that will match each document's fields of expertise. As an example, a perfect algorithm might generate the rankings $D_1 \mapsto C_3 C_4 C_5 C_1 C_2$ and $D_6 \mapsto C_4 C_5 C_3 C_2 C_1$.}
\label{fig:data}
\end{figure}
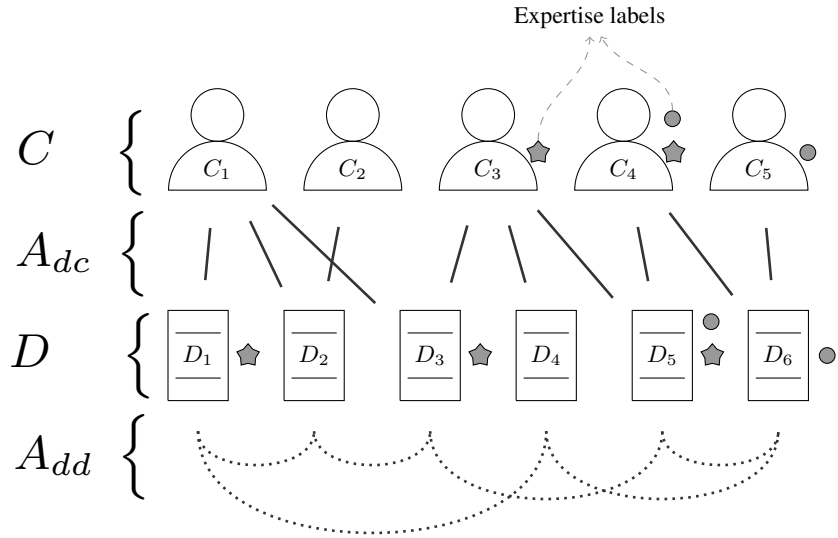 

To evaluate the candidate scores provided by the algorithms, we compare the resulting rankings with the ground truth fields of expertise. If a document is associated with three different labels, we expect the algorithm to rank first all experts associated to at least one of these labels. We report the area under the ROC curve (AUC), the precision at 10 (P@10) and the average precision (AP) and we compute their standard deviation along the queries. That is, we evaluate the robustness of the algorithms against the variety of document-queries. 

\subsection{Datasets}

We consider 4 datasets. The first one is an extract of DBLP \cite{tang2008arnetminer} in which a list 199 experts in 7 fields are annotated \cite{zhang2007expert} by human judgments \footnote{\url{https://lfs.aminer.cn/lab-datasets/expertfinding/##expert-list}}. Our dataset only considers the annotated experts and the other candidates that are close in the co-authorship network which explains the relatively small size of our network compared to the original one. In addition to the expert annotations, our evaluation framework requires document annotations since we adopt the document-query methodology for expertise retrieval. We asked two PhD students in computer science to associate independently 20 randomly drawn documents per field of expertise (140 in total). Then, only the labels on which the two annotators agreed were kept, leaving 114 annotated papers. The mean Cohen's kappa coefficient across the labels is $0.718$. An advantage of our methodology is that we can evaluate the algorithms on more queries (114 documents) than the traditional method (7 labels). This allows us to assess the robustness of the algorithms by computing the standard deviations of the ranking metrics along all queries. However, one might suggest that these 7 labels do not reflect a representative set of expertise as there are too broad. For this reason, we seek for a wider granularity of expertise by the use of well-know \textit{question \& answer} website.   

If scientific publication networks are easy to find on the Web, scientific expertise annotations are rarely available for both authors and publications. We use data downloaded in June 2019 from Stack Exchange \footnote{\url{https://archive.org/details/stackexchange}} to create datasets for expert finding collected from three communities closely related to research. Academia \footnote{\url{https://academia.stackexchange.com/}} is dedicated to academics and higher education. Mathoverflow \footnote{\url{https://mathoverflow.net/}} gathers professional mathematicians and is widely used by researchers. Stats \footnote{\url{https://stats.stackexchange.com/}} (also known as \textit{Cross Validated}) addresses statistics, machine learning and data mining issues. 
For each dataset, we first keep questions with at least 10 user votes that have at least one answer with 10 user votes or more. We build the networks by linking questions with their answers and by linking answers with the users who published them. The field of expertise are the tags associated with the questions. Only the tags that occur at least 50 times are kept. We annotate an expert with the tags of a question if her answer to that question received at least 10 votes. Note that the tags are first provided by the users who ask the questions but they are thereafter verified by experimented users.

The general properties of our 4 datasets are presented in Table \ref{tab:datasets}. The annotations and the preprocessed datasets are made publicly available.            

\begin{table}[ht]
\center
\caption{General properties of the datasets.}
\begin{tabular}{l|cccccccc}
                   & \# candidates  & \# documents  & \# labels  & \# experts   & \#  queries  & label example        \\ \hline
DBLP               & 707            & 1641          & 7                      & 199          & 114          & 'information\_extraction'\\
Stats              & 5765           & 14834         & 59                     & 5765         & 3966         & 'maximum-likelihood'     \\
Academia           & 6030           & 20799         & 55                     & 6030         & 4214         & 'recommendation-letter'  \\
Mathoverflow       & 7382           & 38532         & 98                     & 7382         & 10614        & 'galois-representations' \\
\end{tabular}
\label{tab:datasets}
\end{table}

\subsection{Algorithms}

We run the experiments with 4 baseline algorithms and 4 document network embedding algorithms. The laters are adapted with two aggregation schemes in order to deal with the candidates since they are primarily designed for document network only. These aggregations are arbitrary and are voluntarily the most straightforward way to run DNE algorithms on bipartite networks of authors-documents. We further discuss these choices in section \ref{sec:xp}. 

\subsubsection{Baselines}

We run the experiments with the same models as in \cite{brochier2018impact}, using the tf-idf representations and the cosine similarity measure. Also, we add a random model to have reference metrics:

\begin{itemize}
    \item \textbf{Random model}: we randomly draw scores between 0 and 1 for each candidate; 
    \item \textbf{P@noptic model} \cite{craswell2001p}: we concatenate the textual content of each document associated to the candidates, use their tf-idf representations and compute the cosine similarity to produce the scores;
    \item \textbf{Voting model} \cite{macdonald2006voting}: we use the reciprocal rank to aggregate the scores at the candidate level;
    \item \textbf{Propagation model} \cite{serdyukov2008modeling}: we concatenate the two adjacency matrices $A_{dc}$ and $A_{dd}$ to construct a transition matrix between candidates and documents such that \linebreak $A = \left(\begin{array}{cc} A_{dd} & A_{dc} \\ A_{dc}^{\intercal} & 0 \end{array}\right)$. The initial scores are the cosine similarities between the tf-idf representations of the query and the documents. The scores are propagated iteratively until convergence with a restart probability of $0.5$.  
\end{itemize} 

We also run the voting and propagation models using document representations produced by IDNE in place of the tf-idf vectors. The document network provided to IDNE has adjacency matrix $A_d = A_{dc} A_{dc}^{\intercal} + A_{dd}$. 

\subsubsection{Extending DNE algorithms for expert finding}

DNE methods usually operate in networks of documents, with no candidate nodes. To apply them in the context of expert finding, we propose two straightforward approaches:

\begin{itemize}
    \item \textbf{pre-aggregation}: as in the P@noptic model, meta-documents are generated by aggregating the documents produced by each candidates. Furthermore, an adjacency matrix of a meta-network between candidates and documents is constructed. We compute a candidate network as $A_c = A_{dc}^{\intercal} A_{dc}$ and a document network as $A_d = A_{dc} A_{dc}^{\intercal} + A_{dd}$. The meta-network is then $A = \left(\begin{array}{cc} A_d & A_{dc} \\ A_{dc}^{\intercal} & A_c \end{array}\right)$. The candidate and document representations are then generated by treating this meta-network and the concatenation of the documents and meta-documents as an ordinary instance of document network. From this meta-network, we generate representations with the DNE algorithms. The scores of the candidates are generated by cosine similarity between the representation of the document-query and the representations of the candidates;
    \item \textbf{post-aggregation}: in this setting, we first train the DNE algorithm on the network of documents defined by $A_d = A_{dc} A_{dc}^{\intercal} + A_{dd}$. Once the representations are generated for all documents, a representation for a candidate is computed by averaging the vectors of all documents associated to her. The scores are then computed by cosine similarity. 
\end{itemize}

We run the experiments with 4 document network embedding algorithms, using the authors' implementations. For all methods, the dimension of the representations is set to 256:
\begin{itemize}
    \item \textbf{TADW} \cite{yang2015network}: we follow the original paper by using 20 iterations and a penalty term $\lambda=0.2$;
    \item \textbf{GVNR-t} \cite{brochier2019global}: we use $\gamma=10$ random walks of length $t=40$, a sliding window of size $l=5$ and a threshold $x_{\text{min}}=2$ with 4 iterations;
    \item \textbf{Graph2gauss} (G2G) \cite{bojchevski2018deep}: we make sure the loss function converges before the maximum number of iterations;
    \item \textbf{IDNE} \cite{brochier2020inductive}: we run all experiments with $n_t=32$ topic vectors with 5000 balanced mini-batches of 16 positive samples and 16 negative samples.
\end{itemize}

\section{Experiment Results} \label{sec:xp}

Tables \ref{table:dblp} to \ref{table:mathoverflow} show the experiments results. In the following, we analyze the performances of the aggregation scheme against the baseline algorithms, we highlight the interesting results obtained when using the baselines with pre-computed document representations with a DNE algorithm and finally we make some observations on the differences between the datasets. Note that the implementation of TADW, provided by the authors, could not scale to Mathoverflow.  
 
\subsection{Baselines versus DNE aggregation schemes}

For all datasets, the propagation model performs generally better than the other algorithms, particularly in terms of precision. Both aggregation schemes yield to poor results and none of these two methods appear to be better than the other. GVNR-t is the best algorithm among the document network embedding models. We believe that, if DNE algorithms are well suited for document network representation learning, the gap between simple tasks such as node classification and link prediction and the task of expert finding is too big for our naive aggregation schemes to perform well. Especially, the network structure changes significantly between an homogeneous network and an heterogeneous network. Moreover, expert finding algorithms often benefit from information about the centrality of the candidates and documents. DNE algorithms do not particularly preserve this information neither do our aggregation schemes.       

\subsection{Using DNE as document representations for the baselines}

\begin{figure}[t!]
\begin{subfigure}[b]{0.48\textwidth}
    \centering
    \includegraphics[scale=0.4]{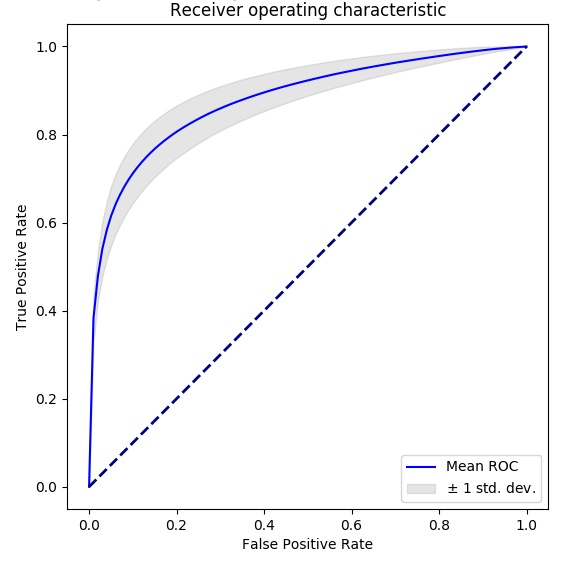}
    \caption{Propagation model with tf-idf representations: the curve has a nice shape which means the ranking of candidates are good even for the last ranked experts.}
   \label{fig:prop_tfidf}
\end{subfigure}%
~~~~
\begin{subfigure}[b]{0.48\textwidth}
    \centering
    \includegraphics[scale=0.4]{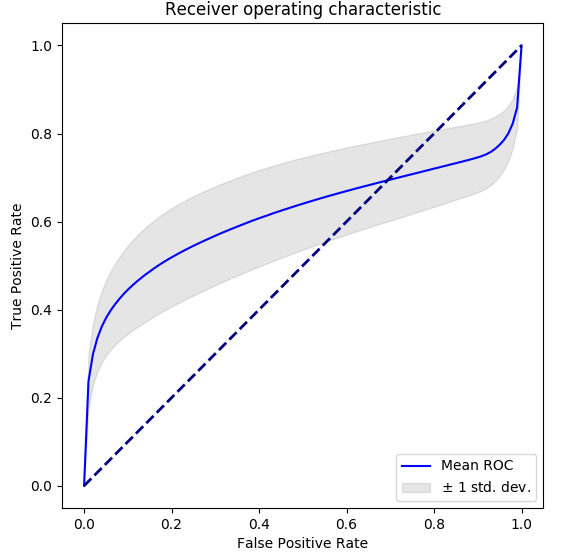}
    \caption{Propagation model with IDNE representations: the first ranked candidates are good but the algorithm tends to wrongly rank last many true experts.}
   \label{fig:prop_tfidf}
\end{subfigure}
\caption{Effect of IDNE representations on the propagation model. Using document network embeddings significantly damages the rankings.}
\label{fig:auc}
\end{figure}

Since the baseline algorithms perform well, we study the possibility to apply them using a DNE algorithm for the representations of the documents. We only report the results with the representations computed with IDNE but we observe the same behaviors with other DNE models. First, these representations constantly improve the voting model, which achieves best results in terms of AUC on Stats and Mathoverflow. Then, the most surprising effect is the significant decrease of performance of the propagation model. If the precision for the first ranked candidates is not affected, the AUC score significantly drops for the three \textit{Q\&{}A} datasets. We believe that document network embeddings captures too long-range dependencies between the documents in the network, which are then subsequently exaggerated by the propagation procedure. Figure \ref{fig:auc} shows the effect of the representations used with the propagation model on the ROC curve.  

\subsection{Differences between the datasets}

The results achieved by the algorithms on all three Stack Exchange datasets are consistent. However, they do not behave the same with DBLP. First, DNE methods get closer scores to the baselines on DBLP. In the \textit{Q\&A} datasets, the interactions are more isolated i.e. there are more users having fewer interactions. This difference of network properties might disadvantage DNE methods who are usually trained on scale-free networks whose degree distribution follows a power law. Moreover, the propagation method does not suffer with DBLP from the decrease of performance induced by the IDNE representations. We hypothesize that the low number of expertise fields associated with this dataset largely reduces the effect described in the previous section.

\section{Discussion and Future Work} \label{sec:conclu}

In this paper, we provide experiment materials for expert finding with the help of four annotated datasets and further report results based on several baseline algorithms. Moreover, we study the ability of document network embedding methods to tackle the expert finding challenge. We show that DNE algorithms can not be trivially adapted to achieve state-of-the-art scores. However, we reveal that document network embeddings can improve the voting model but diminish the propagation model. 

In future work, we would like to find an efficient way to bridge the gap between DNE algorithms and expert finding. To do so, taking the heterogeneity into account should help better capturing the real similarity between a document and a candidate. Furthermore, a deeper analysis of the interplay between the candidates and the text content of the documents appears to be a necessary way to better understand the task of expert finding.  

\clearpage
\newpage

\setlength{\tabcolsep}{4pt}

\begin{table}[H]
\center
\caption{Mean scores with their standard deviations on DBLP}
\label{table:dblp}
\begin{tabular}{lccc}
 & AUC        & P@10       & AP                      \\ \hline
random  & 49.47 (09.80) & 05.00 (06.66) & 07.09 (03.81)             \\
panoptic (tf-idf)  & 74.06(12.94) & 22.37 (16.35) & 23.24 (12.55)             \\
voting (tf-idf)  & 78.60 (11.97) & 26.05 (15.76) & 28.24 (13.92)             \\
propagation (tf-idf) & 79.26 (13.09) & 33.07 (19.61) & 34.66 (18.21)             \\ \hline
pre-agg TADW & 65.84 (12.94) & 15.61 (11.63) & 17.26 (08.78)             \\
pre-agg GVNR-t & 76.90 (11.46) & 19.04 (11.70) & 21.39 (09.61)             \\
pre-agg G2G & 72.87 (12.75) & 15.70 (11.62) & 18.53 (09.37)             \\
pre-agg IDNE & 78.08 (11.27) & 20.18 (11.85) & 22.00 (09.87)             \\ \hline
post-agg TADW & 68.01 (13.37) & 16.32 (11.57) & 18.01 (08.97           \\ 
post-agg GVNR-t & 73.91 (13.93) & 18.86 (12.19) & 20.57 (10.33)             \\
post-agg G2G & 68.94 (15.23) & 16.23 (12.02) & 18.21 (09.76)             \\
post-agg IDNE & 76.87 (13.36) & 19.04 (14.57) & 21.57 (10.96)             \\ \hline
voting (IDNE) & 82.23 (11.08) & 34.82 (18.46) & 37.27 (16.16)             \\
propagation (IDNE) & \textbf{82.44} (16.14) & \textbf{44.47} (22.91) & \textbf{47.01} (22.06)             \\
\end{tabular}
\end{table}

\begin{table}[H]
\center
\caption{Mean scores with standard deviations on Stats}
\label{table:stats}
\begin{tabular}{lccc}
                       &  AUC      & P@10         & AP                      \\ \hline
random & 50.01 (02.24) & 04.52 (07.02) & 04.96 (02.81)             \\
panoptic (tf-idf) & 79.47 (06.22) & 13.45 (13.39) & 15.22 (05.62)             \\
voting (tf-idf)  & 84.96 (05.22) & 52.53 (16.13) & 31.01 (06.58)             \\ 
propagation (tf-idf) & 86.33 (05.64) & \textbf{91.53} (13.44) & \textbf{44.09} (07.70            \\ \hline
pre-agg TADW & 63.07 (07.70) & 11.42 (12.34) & 08.45 (03.87)             \\
pre-agg GVNR-t & 70.67 (09.49) & 21.12 (20.99) & 12.43 (07.30            \\ 
pre-agg G2G & 63.63 (07.62) & 12.93 (12.06) & 07.81 (04.15            \\
pre-agg IDNE & 65.07 (09.05) & 13.37 (13.48) & 09.40 (05.19)             \\ \hline
post-agg TADW & 68.74 (07.02) & 13.67 (12.59) & 09.99 (04.37)             \\
post-agg GVNR-t & 66.56 (08.61) & 22.47 (15.92) & 10.75 (05.42         \\
post-agg G2G  & 62.53 (07.44) & 11.95 (11.86) & 07.48 (04.13)             \\
post-agg IDNE & 65.63 (08.57) & 13.34 (13.13) & 09.38 (04.94            \\ \hline
voting (IDNE) & \textbf{86.94} (04.91) & 53.91 (18.06) & 32.18 (08.33)             \\
propagation (IDNE) & 67.62 (10.11) & 90.43 (15.20) & 33.07 (08.93)             \\
\end{tabular}
\end{table}

\clearpage
\newpage

\begin{table}[H]
\center
\caption{Mean scores with standard deviations on Academia}
\label{table:academia}
\begin{tabular}{lccc}
                          &  AUC      & P@10         & AP                      \\ \hline
random  & 50.02 (01.78) & 05.93 (08.07) & 06.09 (02.72)             \\
panoptic (tf-idf) & 81.54 (04.36) & 18.35 (18.76) & 22.93 (07.14)             \\
voting (tf-idf) & 85.88 (03.47) & 57.99 (15.87) & 37.66 (05.83           \\
propagation (tf-idf) & \textbf{88.02} (03.32) & \textbf{99.01} (03.57) & \textbf{54.04} (05.44)             \\ \hline
pre-agg TADW & 61.47 (06.16) & 11.09 (12.04) & 09.29 (03.53)             \\
pre-agg GVNR-t & 64.22 (09.69) & 25.67 (23.27) & 13.07 (07.54)             \\
pre-agg G2G & 61.54 (05.38) & 14.30 (12.91) & 08.74 (03.69)             \\
pre-agg IDNE & 58.74 (07.49) & 10.21 (11.58) & 08.41 (03.99            \\ \hline
post-agg TADW & 71.94 (04.63) & 14.44 (12.87) & 12.68 (04.37         \\
post-agg GVNR-t & 61.22 (06.24) & 20.70 (14.59) & 10.19 (04.21)             \\
post-agg G2G& 58.87 (05.79) & 12.80 (12.06) & 08.12 (03.67)             \\
post-agg IDNE& 59.97 (07.40) & 10.61 (11.19) & 08.76 (04.17))             \\ \hline
voting (IDNE) & 86.79 (03.90) & 55.81 (17.35) & 37.13 (07.58)             \\
propagation (IDNE)& 61.35 (08.56) & 95.02 (10.15) & 31.27 (08.21)             \\
\end{tabular}
\end{table}

\begin{table}[H]
\center
\caption{Mean scores with standard deviations on Mathoverflow}
\label{table:mathoverflow}
\begin{tabular}{lccc}
                         &     AUC    & P@10       & AP                                 \\ \hline
random& 49.98 (01.62) & 06.44 (08.28) & 06.53 (03.06)             \\
panoptic (tf-idf)& 81.87 (04.46) & 21.95 (19.15) & 22.95 (07.54)             \\
voting (tf-idf)& 86.80 (03.23) & 61.11 (18.68) & 40.10 (08.27)             \\ 
propagation (tf-idf)& 88.08 (03.38) & \textbf{93.68} (12.16) & \textbf{49.58} (08.90)             \\ \hline
pre-agg TADW& NA           & NA           & NA                              \\
pre-agg GVNR-t & 65.34 (09.22) & 44.02 (28.31) & 16.88 (08.55)             \\
pre-agg G2G & 66.84 (08.99) & 22.95 (17.81) & 12.49 (05.70)             \\
pre-agg IDNE & 67.01 (09.26) & 22.96 (17.84) & 13.40 (06.02)             \\ \hline
post-agg TADW  & NA       & NA   & NA                            \\
post-agg GVNR-t & 63.84 (07.59) & 41.81 (22.68) & 14.96 (06.25)             \\
post-agg G2G  & 65.06 (09.09) & 22.43 (16.94) & 11.78 (05.51)             \\
post-agg IDNE & 66.74 (09.10) & 21.92 (17.21) & 13.11 (05.87)             \\ \hline
voting (IDNE) & \textbf{88.71} (03.76) & 68.46 (18.53) & 43.53 (09.90)             \\
propagation (IDNE) & 69.38 (09.65) & 92.35 (13.88) & 39.62 (09.89)             \\
\end{tabular}
\end{table}

\clearpage
\newpage

\bibliography{bibli}

\end{document}